\documentclass[pra]{revtex4}
\usepackage{amsmath,amssymb,mathrsfs}
\usepackage{psfrag}
\usepackage{graphicx}
\usepackage{graphics}
\usepackage{epsfig}
\usepackage{bm}
\usepackage{color}
\usepackage{verbatim,color,ulem}

\newcommand{\beq}{\begin{equation}}
\newcommand{\eeq}{\end{equation}}
\newcommand{\beqa}{\begin{eqnarray}}
\newcommand{\eeqa}{\end{eqnarray}}

\newcommand{\cblue}{\color{black}}

\begin{document}

\title{Density of states for the Unitary Fermi gas 
and the Schwarzschild black hole}

\author{Luca Salasnich$^{1,2,3}$}

\affiliation{$^{1}$Dipartimento di Fisica e Astronomia 
``Galileo Galilei'' and Padua QTech, 
Universit\`a di Padova, Via Marzolo 8, 35131 Padova, Italy \\
$^{2}$Istituto Nazionale di Ottica (INO) del Consiglio Nazionale 
delle Ricerche (CNR), Via Nello Carrara 1, 50019 Sesto Fiorentino, Italy\\
$^{3}$Istituto Nazionale di Fisica Nucleare (INFN), Sezione di Padova, 
Via Marzolo 8, 35131 Padova, Italy}

\begin{abstract}
The density of states of a quantum system 
can be calculated from its definition but, in some cases,  
this approach is quite cumbersome. Alternatively, 
the density of states can be deduced from 
the microcanonical entropy or from the canonical partition function. 
After {\cblue discussing}  the relationship among these procedures, 
we suggest a simple numerical method, which is equivalent 
in the thermodynamic limit to perform a Legendre transformation, 
to obtain the density of states 
from the Helmholtz free energy. We apply this method to determine the 
many-body density of states of the unitary Fermi gas, 
a very dilute system of identical fermions interacting with 
divergent scattering length. {\cblue The unitary Fermi gas is 
highy symmetric due to the absence of any internal scale 
except for the average distance between two particles and, for this reason, 
its equation of state is called universal.} 
{\cblue In the last part of the paper}, 
by using the same {\cblue thermodynamical} techniques,  
we {\cblue review some properties of} the density of states 
of a Schwarzschild black hole, 
which shares with the unitary Fermi gas the problem of finding 
the density of states directly from its definition.
\end{abstract}

\maketitle

\section{Introduction}

The density of states appears in many contexts of statistical mechanics 
\cite{huang} and quantum physics \cite{landau-book}. 
{\cblue The density of states, which tells you how many quantum states exist 
in a given range of energy (or momentum), is extremely useful 
in the experimental and theoretical determination of several physical 
quantities \cite{huang,landau-book}.} In some cases one 
deals with the single-particle density of states, namely the density of states 
of a single quantum particle in the presence of an external potential. 
The determination of this single-particle density of states 
is often quite simple. Instead, the calculation of the many-body density 
of states, i.e. the density of states of system composed by many 
interacting quantum particles, is usually a difficult task. 
Indeed, although the density of states of a quantum system can be, 
in principle, derived from its definition, this approach is 
not always straightforward, in particular for many-body problems. 
As an alternative, the density of states can be deduced from 
the microcanonical entropy or the canonical partition function. 
As well know, in the appropriate thermodynamic limit, 
microcanonical observables can be related to the 
corresponding canonical ones by means of a Legendre 
transformation \cite{huang}. 

In this paper we suggest a straightforward technique for deriving 
the density of states from the Helmholtz free energy. 
This procedure {\cblue is} nothing else than a Legendre transformation 
of the entropy from the canonical ensemble to the microcanonical ensemble. 
We apply this method to calculate the many-body 
density of states of the unitary Fermi gas, characterized by an interaction 
potential with a divergent s-wave scattering length \cite{zwerger2011}. 
When the s-wave scattering length becomes very large the attractive Fermi gas 
of fermionic pairs with opposite spins is made of weakly bound dimers. 
Strictly speaking, the unitary Fermi gas is made of dimers with 
zero binding energy \cite{zwerger2011}. {\cblue This system is 
very peculiar due to the absence of any intrinsic parameter, 
except for the number density. As a consequence, its equation 
of state is called universal \cite{zwerger2011}. 
In addition, for the unitary Fermi gas, 
the conformal invariance plays an important role, as discussed by 
Son and Wingate \cite{son2006}.}  
{\cblue We show that a direct microcanonical evaluation 
of the many-body density of states of the unitary Fermi gas 
gives rise to a formula which seems intractable. 
Then, we exhibit an elegant derivation 
of the many-body density of states of the unitary Fermi gas 
starting from the canonical ensemble and applying a Legendre 
transformation. We believe that this canonical approach could be applied 
to other many-body systems, for instance atomic nuclei, 
Bose-Einstein condensates, and superconductors.} 

Taking into account that the density of states 
is currently an very hot topic in the physics of black holes 
also for the condensed matter theoreticians \cite{sachdev2022}, 
{\cblue for the sake of entertainment}, in the last part of the paper 
we {\cblue review} the density of states 
and other thermodynamical quantities of the Schwarzschild black hole. 
{\cblue Also for this astrophysical system we derive the density of states 
starting from the canonical ensemble 
and applying a Legendre transformation. In this case, however,  
the procedure is quite simple}.

\section{General properties of the density of states}

Let us consider a quantum system with microscopic Hamiltonian ${\hat H}$ and 
macroscopic internal energy between $E$ and $E+{\Delta}$, 
with ${\Delta} \ll E$ \cite{huang}. Following the 
Boltzmann's idea \cite{cercignani}, in the microcanonical ensemble 
the entropy $S(E)$ of the system can be written as 
\beq 
S(E) = k_B \ln(W(E)) \; ,
\eeq
where $k_B$ is the Boltzmann constant and $W(E)$ is the number of accessible 
microstates between $E$ and $E+\Delta$, that we call adimensional 
density of states, given by \cite{huang}
\beq 
W(E) = N(E+\Delta) - N(E) \; , 
\eeq
where 
\beq 
N(E) = \mathrm{Tr}[\Theta(E-{\hat H})] 
\eeq
is the number of states up to the energy $E$, with $\Theta(x)$ the Heaviside 
step function \cite{huang}. If $\Delta$ is sufficiently small one has 
\beq 
W(E) \simeq D(E) \ {\Delta} \; ,  
\eeq
where 
\beq 
D(E) = \mathrm{Tr}[\delta(E-{\hat H})] 
\label{dosso}
\eeq
is the density of states, with $\mathrm{Tr}$ the trace on the Hilbert space 
of quantum states and $\delta(x)$ the Dirac delta function \cite{huang}. 
In the thermodynamic limit one often writes 
\beq
S(E) \simeq k_B \ln(D(E) E_s) \;  , 
\eeq
with $E_s$ a characteristic energy scale of the system (for instance 
$E_s=\hbar^2n^{2/3}/m$ for $N$ identical particles of mass $m$ 
in a volume $V$ and number density $n=N/V$), because the intensive quantity 
$\ln(\Delta/E_s)$ becomes negligible with respect 
to the extensive quantity $\ln(D(E)E_s)$ \cite{huang}. 

Knowing the Hamiltonian ${\hat H}$ one can calculate $D(E)$ by using 
Eq. (\ref{dosso}). Alternatively, knowing the microcanical 
entropy $S(E)$, one easily derives the adimensional density of states $W(E)$ 
from the entropy $S(E)$ as 
\beq 
W(E) = e^{S(E)/k_B} \; . 
\label{dovvio}
\eeq
The third principle of thermodynamics \cite{huang} states that 
$S(E_{gs})=0$ with $E_{gs}$ the ground-state energy of the system. 
Consequently, from Eq. (\ref{dovvio}) we obtain $W(E_{gs}) =1$. 

In the canonical ensemble, the Helmholtz free energy $F(T)$ of the system 
at temperature $T$ is given by \cite{huang}
\beq 
F(T) = - k_B T \ \ln{({\mathcal Z}(T))} \; , 
\label{fvst}
\eeq
where ${\mathcal Z}(T)$ is the partition function, defined as 
\beq 
{\mathcal Z}(T) = \mathrm{Tr} [ e^{-{\hat H}/(k_BT)}] \; . 
\label{zvst}
\eeq 
It is not difficult to show that the partition function ${\mathcal Z}(T)$ 
is directly related to the density of states $D(E)$. In fact, 
\beq 
\mathrm{Tr}[ e^{-{\hat H}/(k_BT)}] = \mathrm{Tr}[ 
\int dE \ \delta(E - {\hat H}) \ e^{-E/(k_BT)}] 
= \int dE \ \mathrm{Tr}[ \delta(E - {\hat H})] \ e^{-E/(k_BT)}
\eeq
and consequently 
\beq 
{\mathcal Z}(T) = \int dE \ D(E) \ e^{-E/(k_BT)} \; .
\eeq
Inverting this formula one gets the density of states $D(E)$ as 
a function of the partition function ${\mathcal Z}(T)$, and then also 
$D(E)$ as a function of the Helmholtz free energy $F(T)$. However, 
this procedure is quite cumbersome because it involves the calculation 
of an anti-Laplace transformation. 

In this paper we suggest a much simpler procedure to obtain 
the adimensional density of states $W(E)$ from the Helmholtz 
free energy $F(T)$. 
In the canonical ensemble, the entropy $S$ as a function of the 
temperature $T$, namely $S(T)$, is given by 
\beq 
S(T) = - \left({\partial F(T)\over \partial T}\right)_{N,V} \; , 
\label{svst}
\eeq
that is the partial derivative of the  Helmholtz free energy $F(T)$ 
with respect to the temperature $T$ at fixed number $N$ of particles and 
volume $V$. Moreover, the internal energy $E(T)$ reads 
\beq 
E(T) = F(T) + T \ S(T) \; . 
\label{evst}
\eeq
Both $S(T)$ and $E(T)$ {\cblue depend on} the temperature $T$. 
This means that $T$ can be considered as a dummy variable to get, 
or analytically or numerically, the parametric curve $S$ vs $E$, 
i.e. $S=S(E)$, which could be a multivalued function. 
In this way we are actually performing, in the thermodynamic limit, 
a Legendre transformation of the entropy from the canonical ensemble to the 
microcanonical ensemble. Having this result, one can then 
use Eq. (\ref{dovvio}) to find the adimensional density of states $W(E)$. 

\section{Unitary Fermi gas}

In 2004 the crossover from the Bardeen-Cooper-Schrieffer (BCS) state 
of weakly-correlated pairs of fermions to the Bose-Einstein condensation 
(BEC) of diatomic molecules was observed with ultracold 
gases of two-component fermionic $^{40}$K or $^6$Li 
atoms \cite{regal2004,zw2004,kinast2004}. 
This BCS-BEC crossover is obtained by using a Fano-Feshbach resonance to change 
the strength of the effective inter-atomic attraction and, 
consequently, the 3D s-wave scattering length $a$ 
\cite{giorgini2008,zwerger2011}. 

Given a gas of $N$ atomic fermions in a volume $V$
with two equally-populated spin components, i.e. 
$N_{\uparrow}=N_{\downarrow}=N/2$, the system is dilute if the characteristic range 
$r_e$ of the inter-atomic potential is much smaller than the average 
interparticle separation $d=n^{-1/3}$ with $n=N/V$ 
the total number density, namely 
\beq 
r_e \ll d \; . 
\eeq
The system is strongly-interacting if the scattering length 
$a$ of the inter-atomic potential greatly exceeds 
the average interparticle separation $d=n^{-1/3}$, i.e. 
\beq
d \ll |a| \; . 
\eeq
The unitarity regime \cite{giorgini2008} is characterized by both 
these conditions: 
\beq 
r_e \ll d \ll |a| \; . 
\eeq
Under these conditions the dilute but strongly-interacting Fermi gas 
is called unitary Fermi gas. 

Ideally, the unitarity limit corresponds to
\beq 
r_e = 0 \quad\quad\quad \mbox{and}\quad\quad\quad a=\pm \infty \; . 
\eeq
In a uniform configuration and at zero temperature, 
the only length characterizing the Fermi gas in the unitarity limit is the
average interparticle distance $d=n^{-1/3}$. 

In this case, simply for dimensional reasons, 
the ground-state energy must be \cite{zwerger2011} 
\beq 
E_{gs} = { \xi} {3\over 5} {\hbar^2\over 2m} (3\pi^2)^{2/3} n^{2/3} N = 
{ \xi} {3\over 5} \epsilon_F N  
\eeq
with $\epsilon_F= \hbar^2(3\pi^2)^{2/3} n^{2/3}/(2m)$ 
Fermi energy of the ideal gas and ${ \xi}$ a 
universal unknown parameter: the Bertsch parameter. 
Monte Carlo calculations and experimental data with dilute and ultracold
atoms suggest that, at zero temperature, the unitary Fermi gas is a 
superfuid with ${ \xi} \simeq 0.4$ \cite{zwerger2011}.

We model \cite{magierski2006,sala2010,sala2022} 
the many-body quantum Hamiltonian ${\hat H}$ of the uniform unitary 
Fermi gas with the simple effective Hamiltonian 
\beq 
{\hat H} = E_{gs} + \sum_{\sigma=\uparrow,\downarrow} 
\sum_{\bf k} \epsilon_\text{sp}(k) \ {\hat c}_{{\bf k}\sigma}^{\dagger} 
{\hat c}_{{\bf k}\sigma} + \sum_{\bf q} \epsilon_\text{col}(q) \ 
{\hat b}_{\bf q}^{\dagger} {\hat b}_{\bf q} \; , 
\label{hamilt}
\eeq
where ${\hat c}_{{\bf k}\sigma}^{\dagger}$ is the creation operator 
and ${\hat c}_{{\bf k}\sigma}$ the annihilation operator of fermionic 
single-particle excitations characterized by energy $\epsilon_\text{sp}(k)$, 
spin $\sigma$, and wavevector $\textbf{k}$. 
Similarly, ${\hat b}_{\bf q}^{\dagger}$ is the creation operator and 
${\hat b}_{\bf q}$ the annihilation operator of bosonic collective 
excitations with energy $\epsilon_\text{col}(q)$ and wavevector ${\bf q}$.

The energy of the the BCS-like excitations can be written as 
\beq 
\epsilon_\text{sp}(k) = 
\sqrt{\left({\hbar^2 k^2\over 2m} - \zeta \epsilon_F \right)^2 + \Delta_0^2}  
\eeq
where $\zeta=0.9$ takes into account many-body effects 
on the Fermi surface \cite{magierski2009}. 
Instead, $\Delta_0=\gamma \epsilon_F$ is the energy gap 
with $\gamma = 0.45$ \cite{carlson2005}.

The energy of collective elementary excitations is instead 
assumed to be given by 
\beq
\epsilon_\text{col}(q) = \sqrt{\frac{\hbar^2q^2}{2m} \bigg( 2 m c_B^2 + 
{ \lambda} \frac{\hbar^2q^2}{2m} \bigg) } \; , 
\eeq
where $c_B = \sqrt{{ \xi}/3}\ v_F$ with $v_F=\sqrt{2\epsilon_F/m}$. 
{\cblue In Ref.} \cite{sala2010} we used 
the value ${ \lambda}=0.25$, which is consistent with a macroscopic 
time-dependent nonlinear Schr\"odinger equation 
approach without the inclusion of spurious terms \cite{sala2008}.
In a recent paper \cite{sala2022} we used instead 
${ \lambda}=0.08$, which is the value obtained \cite{sala2015} 
from the beyond-mean-field GPF theory \cite{tempere2012} at unitarity. 

\subsection{Attempt of direct evaluation of the many-body 
density of states} 
\label{hard}

We try to write the density of states $D(E)$ of the unitary Fermi gas 
by using Eq. (\ref{dosso}) with Eq. (\ref{hamilt}). We immediately find 
\beq 
D(E) = \sum_{ \{ n_{{\bf k}\sigma} \} } 
\sum_{ \{ n_{\bf q} \} } 
\delta\Big( E - E_{gs} - \sum_{\sigma=\uparrow,\downarrow} 
\sum_{\bf k} \epsilon_\text{sp}(k) \ n_{{\bf k}\sigma} - 
\sum_{\bf q} \epsilon_\text{col}(q) \ n_{\bf q} \Big) \; , 
\label{nonso}
\eeq
where $n_{{\bf k}\sigma}$ and $n_{\bf q}$ are the occupation numbers 
of single-particle and collective excitations, respectively. 
It is important to remark that Eq. (\ref{nonso}) is the many-body 
density of states of the system and not the much more 
familiar single-particle density of states. 

Taking into account the Fourier representation of the Dirac 
delta function $\delta(x)$ we have 
\beqa 
D(E) &=& \sum_{ \{ n_{{\bf k}\sigma} \} } 
\sum_{ \{ n_{\bf q} \} } {1\over 2\pi}\int d\xi \ 
e^{i\xi \big( E - E_{gs} - \sum_{\sigma=\uparrow,\downarrow} 
\sum_{\bf k} \epsilon_\text{sp}(k) \ n_{{\bf k}\sigma} - 
\sum_{\bf q} \epsilon_\text{col}(q) \ n_{\bf q} \big)} 
\nonumber 
\\
&=& {1\over 2\pi}\int d\xi \ e^{i \xi (E-E_{gs})} \ 
\sum_{ \{ n_{{\bf k}\sigma} \} } 
e^{-i\xi \sum_{\sigma=\uparrow,\downarrow} 
\sum_{\bf k} \epsilon_\text{sp}(k) \ n_{{\bf k}\sigma}} 
\ \sum_{ \{ n_{\bf q} \} } 
e^{-i\xi \sum_{\bf q} \epsilon_\text{col}(q) \ n_{\bf q}} 
\nonumber 
\\
&=& {1\over 2\pi}\int d\xi \ e^{i \xi (E-E_{gs})} \ 
\prod_{\sigma=\uparrow,\downarrow} \prod_{\bf k} 
\sum_{n_{{\bf k}\sigma} = 0,1} 
e^{-i\xi \epsilon_\text{sp}(k) \ n_{{\bf k}\sigma}} 
\ \prod_{\bf q} \sum_{n_{\bf q} = 0}^{+\infty} 
e^{-i\xi \epsilon_\text{col}(q) \ n_{\bf q}} 
\nonumber 
\\
&=& {1\over 2\pi}\int d\xi \ e^{i \xi (E-E_{gs})} \ 
\prod_{\sigma=\uparrow,\downarrow} \prod_{\bf k}  
(1 + e^{-i\xi \epsilon_\text{sp}(k)}) 
\ \prod_{\bf q} {1\over 1 - e^{-i \xi  \epsilon_\text{col}(q)}} \; . 
\label{gulp}
\eeqa
Unfortunately, Eq. (\ref{gulp}) does not help very much to obtain 
a tractable expression of the density of states. For this reason, 
in the next Section, we analyze the same system in 
the canonical ensemble, where we will find a similar, but 
more manageable, formula for the partition function. 
The reason is that working in the canonical ensemble the statistical 
independence of non-interacting macroscopic subsystems 
is ensured by the canonical density operator $e^{-{\hat H}/(k_BT)}$, 
which is an exponential operator \cite{huang}. 

\subsection{Canonical ensemble}

As previously discussed, in the canonical ensemble 
the Helmholtz free energy $F(T)$ of the system 
is obtained from the partition function $\mathcal{Z}(T)$ adopting 
Eqs. (\ref{fvst}) and (\ref{zvst}) \cite{huang}. In particular, 
by using Eqs. (\ref{zvst}) and (\ref{hamilt}) we find 
\beqa
\mathcal{Z}(T) &=& \sum_{ \{ n_{{\bf k}\sigma} \} } 
\sum_{ \{ n_{\bf q} \} } 
e^{-\big(E_{gs} + \sum_{\sigma=\uparrow,\downarrow} 
\sum_{\bf k} \epsilon_\text{sp}(k) \ n_{{\bf k}\sigma} + 
\sum_{\bf q} \epsilon_\text{col}(q) \ n_{\bf q} \big)/(k_BT)}
\nonumber 
\\
&=& e^{-E_{gs}/(k_BT)} 
\sum_{ \{ n_{{\bf k}\sigma} \} } 
e^{-\big(\sum_{\sigma=\uparrow,\downarrow} 
\sum_{\bf k} \epsilon_\text{sp}(k) \ n_{{\bf k}\sigma}\big)/(k_BT)} 
\ \sum_{ \{ n_{\bf q} \} } 
e^{-\big(\sum_{\bf q} \epsilon_\text{col}(q) \ n_{\bf q}\big)/(k_BT)} 
\nonumber 
\\
&=& e^{-E_{gs}/(k_BT)}
\prod_{\sigma=\uparrow,\downarrow} \prod_{\bf k} 
\sum_{n_{{\bf k}\sigma} = 0,1} 
e^{-(\epsilon_\text{sp}(k)/(k_BT)) \ n_{{\bf k}\sigma}} 
\ \prod_{\bf q} \sum_{n_{\bf q} = 0}^{+\infty} 
e^{-(\epsilon_\text{col}(q)/(k_BT)) \ n_{\bf q}} 
\eeqa
Thus, we can write 
\beq 
\mathcal{Z}(T) = \mathcal{Z}_\text{gs}(T) \ \mathcal{Z}_\text{sp}(T) 
\ \mathcal{Z}_\text{col}(T) \; , 
\eeq
where 
\beqa
\mathcal{Z}_\text{gs}(T) &=& e^{-E_\text{gs}/(k_BT)} 
\\
\mathcal{Z}_\text{sp}(T) &=& \prod_{\sigma=\uparrow,\downarrow} \prod_{\bf k}  
(1 + e^{-\epsilon_\text{sp}(k)/(k_BT)}) 
\\
\mathcal{Z}_{col}(T) &=& \prod_{\bf q} 
{1\over 1 - e^{-\epsilon_\text{col}(q)/(k_BT)}} \; . 
\eeqa
With the help of Eq. (\ref{fvst}) the corresponding Helmholtz free energy reads
\beq 
F(T) = F_\text{gs}+F_\text{sp}(T)+F_\text{col}(T) \; , 
\eeq
where $F_\text{gs}=E_\text{gs}$,
\beq
F_{\text{sp}}(T) = - 2k_BT \sum_{\mathbf{k}} 
\ln [ 1 + e^{-\epsilon_{\text{sp}}(k)/(k_BT)} ] \; , 
\eeq
and 
\beq
F_{\text{col}}(T) = - k_B T\sum_{\mathbf{q}} 
\ln [ 1 - e^{-\epsilon_{\text{col}}(q)/(k_BT)} ] \; . 
\eeq

Quite remarkably, the total free energy $F(T)$ can be written \cite{sala2010} 
in a compact form as 
\beq 
F(T) = N \epsilon_F \Phi({T\over T_F}) \; , 
\label{free} 
\eeq 
where $\Phi(x)$ depends on the scaled temperature $x \equiv T/T_F$ only, 
with $T_F=\epsilon_F/k_B$ the Fermi emperature. In particular, we have 
\beqa 
\Phi(x) &=& {3\over 5}{ \xi} - 3 x \int_0^{+\infty} 
\ln{\left[ 1 + e^{-{\tilde \epsilon}_\text{sp}(u)/x}\right]} u^2 \mathrm{d} u
\nonumber
\\
&+& {3\over 2} x \int_0^{+\infty} 
\ln{\left[ 1 - e^{-{\tilde \epsilon}_\text{col}(u)/x} \right]}u^2 
\mathrm{d} u \; . 
\label{free-scaled} 
\eeqa
Here the integrals replace the summations.  
For example, $\sum_{\bf k}\to V\int d^3{\bf k}/(2\pi)^3$. 
Moreover, we set ${\tilde \epsilon}_\text{col}(u)=\sqrt{u^2(4{ \xi}/3 + 
{ \lambda} u^2)}$ and ${\tilde \epsilon}_\text{sp}(u)
=\sqrt{(u^2-\zeta)^2+\gamma^2}$.

We can now calculate the entropy $S(T)$ and the internal energy $E(T)$ 
by using Eqs. (\ref{svst}) and (\ref{evst}). In particular, taking into account 
Eq. (\ref{free-scaled}) we find for the entropy 
\beq 
S(T) = - N k_B \Phi'({T\over T_F}) \; .    
\label{entropy}
\eeq 
where $\Phi'(x)=d\Phi(x)/dx$. Furthermore, for the internal energy $E$ 
we obtain the expression 
\beq 
E(T) = N \epsilon_F 
\left[ \Phi ({T\over T_F}) - {T\over T_F} \ \Phi' ({T\over T_F}) \right] \; .  
\label{internal}
\eeq

\begin{figure}[t]
\centerline{\epsfig{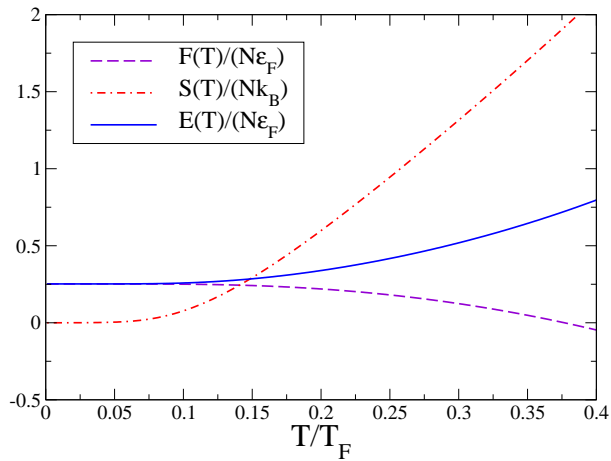}}
\caption{Unitary Fermi gas: 
Scaled free energy $F(T)/(N\epsilon_F)$, scaled entropy $S(T)/(N k_B)$, 
and scaled internal energy $E(T)/(N\epsilon_F)$ deduced from our model, 
as a function of the scaled temperature $T/T_F$ with 
$T_F=\epsilon_F/k_B$ the Fermi temperature.} 
\label{fig1}
\end{figure} 

\begin{figure}[t]
\centerline{\epsfig{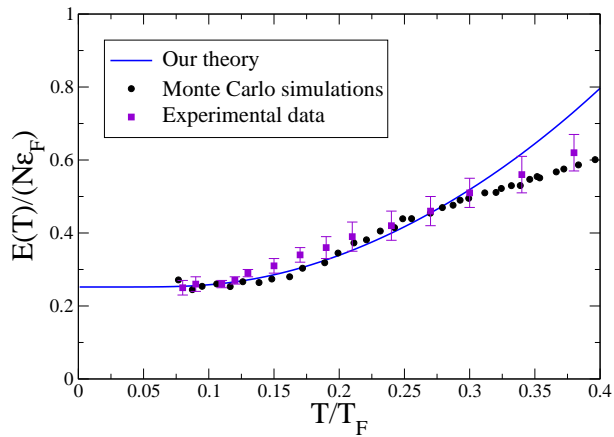}}
\caption{Unitary Fermi gas: Scaled internal energy $E(T)/(N\epsilon_F)$,  
as a function of the scaled temperature $T/T_F$. Solid line is obtained 
with our model. Filled circles: Monte Carlo simulations \cite{bulgac2008}. 
Squares with error bars: experimental data \cite{jap2010}.} 
\label{fig2}
\end{figure} 

In Fig. \ref{fig1} we plot the free energy $F(T)$, the entropy $S(T)$, and 
the internal energy $E(T)$ of the unitary Fermi gas by using Eqs. (\ref{free}), 
(\ref{entropy}), and (\ref{internal}). We choose the following values 
for the parameters of our model: $\xi=0.42$, $\lambda=0.25$, $\zeta=0.9$, 
and $\gamma=0.45$. The figure shows that, by increasing the temperature $T$,  
the Helmholtz free energy $F(T)$ monotonically decreases, while both 
the internal energy $E(T)$ and the entropy $S(T)$ are monotonic growing 
functions. 

It is important to stress that our model for the low-temperature 
thermodyamics of the unitary Fermi gas seems to be in quite good agreement 
with both Monte Carlo simulations \cite{bulgac2008} and experimental data 
\cite{jap2010}. In particular, in Fig. \ref{fig2} we compare our 
internal energy $E(T)$ (solid line) with Monte Carlo calculations (filled 
circles) and experimental results (filled squares). Indeed, the 
agreement among these different datasets is impressive. 
We stress that Eq. (\ref{hamilt}) is a low-temperature Hamiltonian 
because we are not taking into account the fact that, in general, 
the elementary excitations $\epsilon_{sp}(k)$ and $\epsilon_{col}(q)$ 
depend on the temperature $T$. 

\subsection{Numerical calculation of the many-body density of states}
\label{soft}

\begin{figure}[t]
\centerline{\epsfig{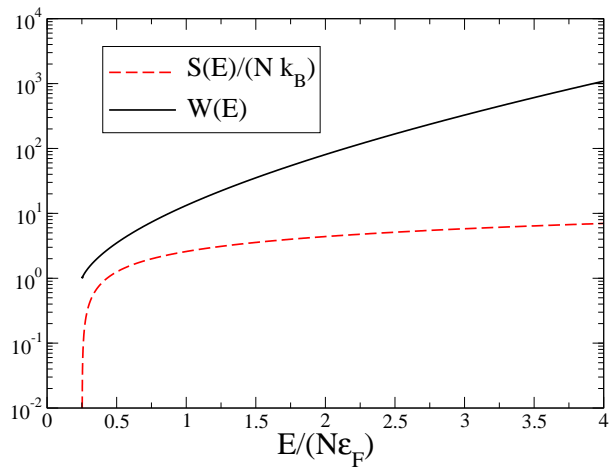}}
\caption{Unitary Fermi gas: The dashed line is the scaled 
entropy $S(E)/(Nk_B)$, as a function of the scaled 
internal energy $E/(N\epsilon_F)$. The solid line is the 
adimensional many-body density of states $W(E)$, as a function of the 
scaled internal energy $E/(N\epsilon_F)$.} 
\label{fig3}
\end{figure}

As previously discussed, having $S(T)$ and $E(T)$, we can immediately 
get the curve $S=S(E)$ by using $T$ as a dummy variable. 
This is a Legendre transformation from $S(T)$ to $S(E)$. 
In Fig. \ref{fig3} we plot this curve (dashed line) and also 
the curve (solid line) of the adimensional many-body density of states 
$W(E)$, which is obtained from Eq. (\ref{dovvio}). 

Fig. \ref{fig3} shows that, by increasing the internal energy $E$, 
there is a monotonic growth of both entropy $S(E)$ and adimensional 
density of states $W(E)$. This is a consequence of the monotonic behavior 
as a function of $T$ of both $S$ and $E$. 
Clearly, $S(E_{gs})=0$, $W(E_{gs})=1$, and  
$W(E)$ is an exponental function of the internal energy $E$. 

\begin{figure}[t]
\centerline{\epsfig{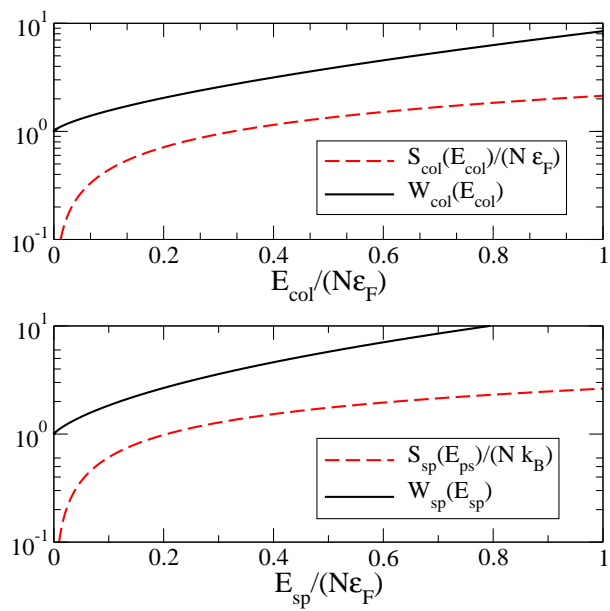}}
\caption{Unitary Fermi gas. Upper panel: The dashed line is the scaled 
entropy $S_{col}(E_{col})/(Nk_B)$ of bosonic collective elementary 
excitations, as a function of 
the scaled internal energy $E_{col}/(N\epsilon_F)$ of the collective elementary 
excitations. The solid line is the adimensional density of 
states $W_{col}(E_{col})$ of collective elementary excitations, 
as a function of the scaled internal energy $E_{col}/(N\epsilon_F)$ 
of collective elementary excitations.
Lower panel: The dashed line is the scaled entropy $S_{sp}(E_{sp})/(Nk_B)$ 
of fermionic single-particle excitations, as a function of 
the scaled internal energy $E_{sp}/(N\epsilon_F)$ of single-particle 
elementary excitations. The solid line is the adimensional density of 
states $W_{sp}(E_{sp})$ of single-particle excitations, as a function of 
the scaled internal energy $E_{sp}/(N\epsilon_F)$ of single-particle 
excitations.} 
\label{fig4}
\end{figure} 

We remark that the entropy (\ref{entropy}) 
and the internal energy (\ref{internal}) are additive, i.e. 
\beqa 
S(T) &=& S_{gs} + S_{sp}(T) + S_{col}(T)  
\\
E(T) &=& E_{gs} + E_{sp}(T) + E_{col}(T) 
\eeqa
with $S_{sg} =0$ and $E_{gs}=(3/5)N\epsilon_F \xi$. 
Moreover, the adimensional density of states $W(E)$
satisfies, in the thermodynamic limit, the equation 
\beq 
W(E) = W_{gs}(E_{gs}) \ W_{sp}(E_{sp}) \ W_{col}(E_{col}) \; , 
\eeq
where $W_{gs}(E_{gs})=e^{S_{gs}(E_{gs})/k_B}=1$, $W_{sp}(E_{sp})=e^{S_{sp}(E_{sp})/k_B}$, 
and $W_{col}(E_{col})=e^{S_{col}(E_{col})/k_B}$. 
In Fig. \ref{fig4} we plot the entropy and the adimensional density 
of states of both collective and single-particle elementary excitations. 

\section{Schwarzschild black hole}

In this section we derive the adimensional density of states 
of a Schwarzschild black hole \cite{sachdev2022,calmet-book} from the 
microcanonical entropy and also from the canonical Helmholtz free energy. 
These are known results but they are however highly non trivial. 
We remark that the derivation of the density of states 
of a black hole directly from its definition is quite controversial 
because a fully consistent quantum Hamiltonian ${\hat H}$ of the 
black hole is not yet available \cite{calmet-book}. 

For a Schwarzschild black hole of mass $M$, 
which does not rotate and has no electric charge, 
the Bekenstein-Hawking entropy \cite{bek1972,haw1975} is given by 
\beq 
S(M) = {4\pi k_B\over \hbar} {G M^2\over c} \; ,  
\eeq
where $G$ is the gravitational constant, $\hbar$ is the reduced Planck constant 
and $c$ is the speed of light in vacuum. Assuming that the internal energy $E$ 
of the system is \cite{ha}
\beq 
E = M c^2 \; , 
\label{mah0}
\eeq
we obtain immediately the microcanonical entropy 
\beq 
S(E) = {4\pi k_B G \over \hbar c^5} E^2 
\label{entroblack}
\eeq
and also, by using Eq. (\ref{dovvio}), the adimensional density of states 
\beq 
W(E) = e^{{4\pi G\over \hbar c^5} E^2} \; . 
\label{mah1}
\eeq
From Eq. (\ref{mah0}) we have $E_{gs}=0$ and, as expected, 
from Eq. (\ref{mah1}) it follows $W(E_{gs})=W(0)=1$. 

\begin{figure}[t]
\centerline{\epsfig{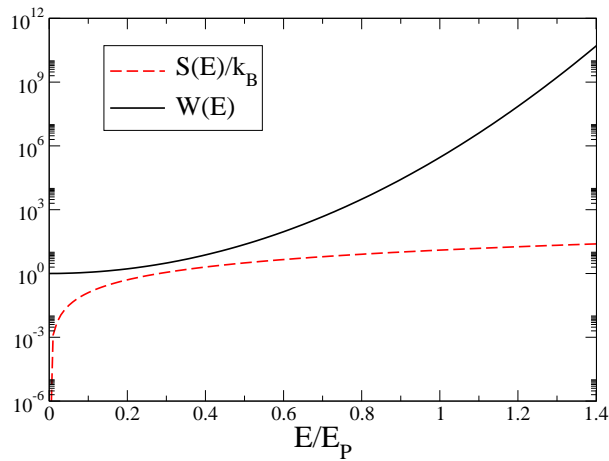}}
\caption{Schwarzschild black hole: The dashed line is the scaled 
entropy $S(E)/k_B$, as a function of the scaled 
internal energy $E/E_P$, with $E_P=\sqrt{\hbar c^5/G}$ the Planck 
energy. The solid line is the adimensional density of states $W(E)$, 
as a function of the scaled internal energy $E/E_P$.} 
\label{fig5}
\end{figure} 

In Fig. \ref{fig5} we plot the curves of the 
scaled entropy $S(E)/k_B$ (dashed line) and of the adimensional 
density of states $W(E)$, obtained with Eqs. (\ref{entroblack}) 
and (\ref{mah1}). The vertical axis is in a log scale to contain the 
huge increase of $W(E)$ with $E$. 

Notice that, in the microcanonical ensemble, 
the temperature $T$ of the system is defined as 
\beq 
{1\over T} = {\partial S(E)\over \partial E} \; . 
\eeq
For the Schwarzschild black hole, using Eq. (\ref{entroblack}) we find 
\beq 
{1\over T} = {8\pi k_B G\over \hbar c^5} E 
\label{equivo}
\eeq
or, equivalently, using also Eq. (\ref{mah0}) we get 
\beq 
T = {\hbar c^3\over 8M\pi k_B G} \; , 
\label{hawtemp}
\eeq
that is the so-called Hawking temperature \cite{haw1974}. 

Let us now consider the Schwarzschild black hole within the framework of the 
canonical ensemble. Because the quantum Hamiltonian ${\hat H}$ of a 
black hole is somehow unknown, Gibbons and Hawking in their approach 
\cite{haw1977} did not use Eq. (\ref{zvst}). Instead, they derived 
the canonical partition function $Z(T)$ of the Schwarzschild black 
hole from the path integral formula 
\beq 
{\mathcal Z}(T) = \int {\cal D}[g_{\mu\nu}(x)] \ e^{-{1\over \hbar} 
\int d^3{\bf x} \int_0^{\hbar/(k_BT)} d\tau \sqrt{g} 
\mathcal{L}(g_{\mu\nu}(x))} \; , 
\label{bohbah} 
\eeq
where $g_{\mu}(x)$ is the metric tensor, $x=(c\tau,{\bf x})$ 
is the space-time coordinate with $\tau$ the imaginary time, 
$g$ is the determinant of the metric tensor, 
and $\mathcal{L}(g_{\mu\nu}(x))$ is the Euclidean Lagrangian density 
of the Einstein-Hilbert action \cite{eh1,eh2}. Taking into account 
the Schwarzschild solution of metric tensor \cite{schw} generated 
by a spherical object of mass $M$ and using a semiclassical approximation 
of Eq. (\ref{bohbah}) with the inclusion of 
appropriate boundary terms, Gibbons and Hawking \cite{haw1977} basically found 
\beq 
{\mathcal Z}(T) = e^{-c^5\hbar/(16\pi G k_B^2T^2)} \; .  
\label{giust}
\eeq
It is important to observe that Eq. (\ref{giust}) was obtained 
by Gibbons and Hawking by using also Eq. (\ref{hawtemp}), 
which is a crucial constraint derived imposing 
the regularity of the Euclidean Schwarzschild metric at the 
Schwarzschild radius $r_s=2 GM/c^2$. 

From Eqs. (\ref{fvst}) and (\ref{giust}) we then get 
the Helmholtz free energy 
\beq 
F(T) = {c^5\hbar\over 16\pi G k_BT} \; . 
\label{fvio}
\eeq
We now calculate the entropy $S(T)$ and the internal energy $E(T)$ 
by using Eqs. (\ref{svst}) and (\ref{evst}). We find 
\beq 
S(T) = {c^5\hbar\over 16\pi G k_BT^2}
\label{svio}
\eeq
and 
\beq 
E(T) = {c^5\hbar\over 8\pi G k_BT} \; .   
\label{evio}
\eeq

\begin{figure}[t]
\centerline{\epsfig{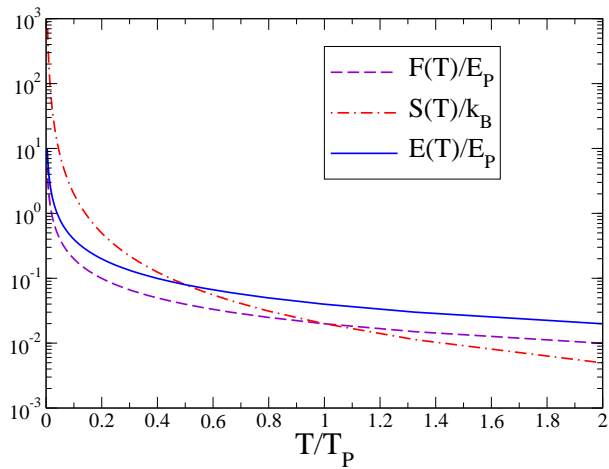}}
\caption{Schwarzschild black hole: 
Scaled free energy $F(T)/E_P$, scaled entropy $S(T)k_B$, 
and scaled internal energy $E/E_P$ 
as a function of the scaled temperature $T/T_P$ with $T_P=E_P/k_B$ 
the Planck temperature and and $E_P=\sqrt{\hbar c^5/G}$ the Planck 
energy.} 
\label{fig6}
\end{figure} 

For the sake of completeness, in Fig. \ref{fig6} we report 
the free energy $F(T)$, the entropy $S(T)$, and 
the internal energy $E(T)$, obtained with Eqs. (\ref{fvio}), 
(\ref{svio}), and (\ref{evio}). 
The figure shows the very unusual behavior of these quantities 
by increasing the temperature $T$: they are all monotonically decreasing. 

As previously discussed, both the canonical entropy $S(T)$ and 
the canonical internal energy $E(T)$ are functions of the 
absolute temperature $T$, which can be considered as a dummy variable to get 
the parametric curve $S$ vs $E$. In this case we can directly find 
the inverse of Eq. (\ref{evio}), which is exactly Eq. (\ref{equivo}). 
Inserting this formula into Eq. (\ref{svio}), i.e. performing 
analytically a Legendre transformation, we obtain the 
microcanonical entropy $S(E)$ of Eq. (\ref{entroblack}) and finally the 
adimensional density of states $W(E)$ given, again, by Eq. (\ref{mah1}). 

\section{Conclusions}

{\cblue In conclusion, we stress again that the knowledge of the number of 
quantum states in a given range of energy is a crucial quantity 
in many context of Physics.} Although {\cblue the} density 
of states can be determined from its {\cblue microcanonical} definition, 
this method is not always simple. As an alternative, 
the density of states can be inferred from the microcanonical entropy 
or the canonical partition. After discussing how these processes 
relate to one another, we have offered a straightforward technique,  
based on the Legendre transformation, for deriving the density 
of states from the Helmholtz free energy. 
As an enlightening example, the unitary Fermi gas, an extremely dilute 
system of identical fermions interacting with divergent scattering length, 
has been studied to determine the many-body density of states. 
In Section \ref{hard} we have found that the computation 
of the density of states from its definition is an hard task, 
while in Section \ref{soft} we have obtained it quite esily 
working in the canonical ensemble. 
Finally, we have used the same thermodynamical framework 
to {\cblue review} the adimensional density of states of a Schwarzschild 
black hole with the Gibbson-Hawking formalism. 
Also in this case, a calculation of the density of states 
from its definition is highly demanding. However, for the 
Schwarzschild black hole, one gets quite easily the density of 
state {\cblue from} the microcanonical entropy 
({\cblue as discussed in several textbooks}) or {\cblue from} 
the canonical free energy.

The author thanks Fulvio Baldovin, {\cblue Davide Cassani}, 
Pieralberto Marchetti, Fabio Sattin, Antonio Trovato, 
and Alexander Yakimenko for useful discussions. 
This work is partially supported by BIRD Project 
"Ultracold atoms in curved geometries" of the University of Padova 
and by Iniziativa Specifica ``Quantum'' of INFN. 
The author acknowledges funding from the European Union-NextGenerationEU, 
within the National Center for HPC, Big Data and Quantum Computing 
(Project No. CN00000013, CN1 Spoke 1: “Quantum Computing”).


\begin{thebibliography}{99}

\bibitem{huang} K-Huang, Statistical Mechanics (Wiley, 2008). 

\bibitem{landau-book} L.D. Landau, E.M. Lifshitz, and L. P. Pitaevskii, 
Course of Theoretical Physics, vol. 9, Statistical Physics: 
Theory of the Condensed State (Butterworth-Heinemann, Oxford, 1980).

\bibitem{zwerger2011} W. Zwerger (Ed.), 
The BCS-BEC Crossover and the Unitary Fermi Gas (Springer, Berlin, 2011). 

{\cblue 
\bibitem{son2006} D.T. Son and M. Wingate, Ann. Phys. {\bf 321}, 197 (2006). 
}

\bibitem{sachdev2022} S. Sachdev, ICTS News {\bf 8}, issue 1 (2022). 

\bibitem{cercignani} C. Cercignani, Ludwig Boltzmann: 
the Man who Trusted Atoms (Oxford Univ. Press, Oxford, 1988). 

\bibitem{giorgini2008} S. Giorgini, L.P. Pitaevskii, and S. Stringari, 
Rev. Mod. Phys. {\bf 80}, 1215 (2008).

\bibitem{regal2004} C.A. Regal et al., Phys. Rev. Lett. 
{\bf 92}, 040403 (2004). 

\bibitem{zw2004} M.W. Zwierlein et al., Phys. Rev. Lett. {\bf 92}, 
120403 (2004). 

\bibitem{kinast2004} J. Kinast et al., Phys. Rev. Lett. {\bf 92}, 
150402 (2004). 

\bibitem{magierski2006} A. Bulgac, J.E. Drut, and P. Magierski, 
Phys. Rev. Lett {\bf 96}, 090404 (2006). 

\bibitem{sala2010} L. Salasnich, Phys. Rev. A {\bf 82}, 063619 (2010). 

\bibitem{sala2022} G. Bighin, A. Cappellaro, and L. Salasnich, 
Phys. Rev. A {\bf 105}, 063329 (2022). 

\bibitem{magierski2009} P. Magierski, G. Wlazlowski, 
A. Bulgac, and J. E. Drut, Phys. Rev. Lett. {\bf 103}, 210403 (2009). 

\bibitem{carlson2005} J. Carlson and S. Reddy, Phys. Rev. Lett. {\bf 95}, 
060401 (2005). 

\bibitem{sala2008} L. Salasnich and F. Toigo, Phys. Rev. A {\bf 78}, 
053626 (2010).

\bibitem{sala2015} G. Bighin, L. Salasnich, P.A. Marchetti, 
and F. Toigo, Phys. Rev. A {\bf 92}, 023638 (2015).

\bibitem{tempere2012} J. Tempere and J. P. Devreese, 
Superconductors: Materials, Properties and Applications, 
InTech {\bf 383} (2012).

\bibitem{bulgac2008} A. Bulgac, J.E. Drut, and P. Magierski, 
Phys. Rev. A {\bf 78}, 023625 (2008). 

\bibitem{jap2010} M. Horikoshi , S. Nakajima, M. Ueda, and T. Mukaiyama, 
Science {\bf 442}, 327 (2010).

\bibitem{calmet-book} X. Calmet (Ed.), Quantum Aspects of Black Holes 
(Springer, 2014). 

\bibitem{bek1972} A. Bekenstein, Lett. Nuovo Cim. {\bf 4}, 99 (1972). 

\bibitem{haw1975} S.W. Hawking, Comm. Math. Phys. {\bf 43}, 199 (1975). 

\bibitem{ha} Y.K. Ha, Gen. Rel. Grav. {\bf 35}, 2045 (2003). 

\bibitem{haw1974} S.W. Hawking, Nature {\bf 248}, 5443 (1974).  

\bibitem{haw1977} G.W. Gibbons and S.W. Hawking, 
Phys. Rev. D {\bf 15}, 2752 (1977).

\bibitem{eh1} S.M. Carroll, Spacetime and Geometry: 
An Introduction to General Relativity (Addison-Wesley, 2004). 

\bibitem{eh2} D. Hilbert, Die Grundlagen der Physik, 
in Nachrichten von der Gesellschaft der Wissenschaften zu Göttingen - 
Mathematisch-Physikalische Klasse {\bf 3}, 395 (1915). 

\bibitem{schw} K. Schwarzschild, 
Sitzungsberichte der Koniglich Preussischen Akademie der Wissenschaften 
zu Berlin, Phys.-Math. Klasse, 189 (1916). 

\end{thebibliography}
\end{document}